\documentclass[twocolumn,preprintnumbers,showpacs,showkeys,amsmath,amssymb,epsf]{revtex4}
\usepackage{graphicx}
\usepackage{dcolumn}
\usepackage{bm}
\begin{document}
\title{A heterogeneous zero-range process related to a two-dimensional walk model}
\author{Seyedeh Raziyeh Masharian}
\email{masharian@iauh.ac.ir}
\affiliation{Islamic Azad University, Hamedan Branch, Hamedan, Iran}
\author{Farhad H. Jafarpour}
\email{farhad@ipm.ir}
\affiliation{Physics Department, Bu-Ali Sina University, 65174-4161 Hamedan, Iran}\date{\today}
\begin{abstract}
We have considered a disordered driven-diffusive system defined on a ring.  This system can be mapped onto a heterogeneous zero-range process. We have shown that the grand-canonical partition function of this process can be obtained using a matrix product formalism and that it is exactly equal to the partition function of a two-dimensional walk model. The canonical partition function of this process is also calculated. Two simple examples are presented in order to confirm the results. 
\end{abstract}
\maketitle
\section{Introduction}
A Zero-Range Process (ZRP), which was first introduced in mathematical literature \cite{S70}, is a widely studied Markov chain in statistical mechanics. A homogenous ZRP is a system of interacting particles which hop between the lattice sites (or boxes) of a homogeneous lattice with the rates that depend only on the number of particles in the departure site. This can be generalized to a Heterogeneous ZRP (HZRP) in which the particles hop between the lattice sites (or boxes) of  a heterogeneous lattice so that the particles' hopping rates also depend on the departure site. Different types of these processes have been introduced and studied in related literature (for a review see \cite{EH05} and references therein). They have also a variety of applications in different fields of physics for instance in transport processes including traffic flows, shaken granular gases, particle condensation and phase separation \cite{KMH05}-\cite{KLMST02}. It is long known and well-stablished that there is possibility for mapping a one-dimensional driven-diffusive system defined on a discrete lattice with periodic boundary conditions onto a ZRP \cite{EH05}. \\
Recently the study of the one-dimensional driven-diffusive systems which can be mapped onto the walk models has drawn physicists attention \cite{BGR04}-\cite{J11}. It has been shown that the simplest system of this type, that is the Asymmetric Simple Exclusion Process (ASEP) with open boundaries, can be mapped onto a walk model consisting of different Dyck paths \cite{BGR04}-\cite{BE04}. The ASEP is a system of hard-core particles moving in a preferred direction on a one-dimensional discrete lattice \cite{DEHP}. The hard-core particles are injected from the left boundary of the lattice and hop exclusively toward the right boundary. Finally they leave the lattice from the right boundary. It has also been shown that the Partially ASEP (PASEP) in which the hard-core particles can hop both to their leftmost and rightmost neighboring empty lattice sites can be mapped onto a weighted walk model \cite{BCEP06}-\cite{BJJK09}.\\
In a recent paper \cite{J11} it has been shown that a driven-diffusive system consisting of two different classes of particles defined on a discrete lattice with periodic boundary condition can be mapped onto a two-dimensional walk model. The system is defined on a ring of length $N$. It consists of $M-1$ first-class particles and a single second-class particle sometimes called the impurity. Both the first and second-class particles hop in the same preferred direction, and there is no overtaking. The steady-state of this system was first studied using a matrix product method (for a review of the matrix product method see \cite{BE07}) in \cite{E96}. It was also shown that the system can be mapped onto a HZRP if one considers the particles (empty lattice sites) in the driven-diffusive system as boxes (particles) in the HZRP \cite{E00}. It turns out that if the single second-class particle moves slow enough, a large number of empty lattice sites pile-up in front of it otherwise the number of empty lattice sites in front of the second-class particle is of order $1$. In the equivalent HZRP these two phases correspond to the cases where the number of particles in the box associated with the second-class particle is either of order of the number of empty lattice sites or $1$. In \cite{J11} the author has shown that the grand-canonical partition function of this disordered driven-diffusive system is equal to that of a walk model defined on the first quarter of a two-dimensional plane. The random walker starts from the origin and takes consecutive steps either upward or downward, provided that it does not lie on the horizontal axis. Whenever the random walker lies on the horizontal axis, it can either goes one step upward or moves horizontally. After taking $N-1$ steps it can get to any height between $0$ and $N-1$. The paths taken by the random walker are weighted. The grand-canonical partition function of walk model is equal to the sum of these weights. If one considers the paths which contain exactly $N-M$ upward steps, then the resulting partition function is exactly that of the disordered driven-diffusive system explained in \cite{E96} and \cite{E00}. \\
In this paper we start with a generalized disordered driven-diffusive system with long-range interactions defined on a one-dimensional discrete lattice with periodic boundary conditions. We show that the steady-state of this system, which is a product measure, can be obtained using a matrix product method. The matrices associated with the particles and the empty lattice sites satisfy an algebra which has not been studied thus far. We discuss how the driven-diffusive system can be mapped onto a HZRP.  Both the canonical and grand-canonical partition functions of the HZRP are calculated. An equivalent walk model is also introduced and studied in detail. Finally, two simple examples are provided to confirm the results.    
\section{The disordered driven-diffusive system}
In \cite{E96} the author has introduced a disorder driven-diffusive system consisting of $M$ different particles which hop in a preferred direction on a closed ring of length $N$. The particles are labeled with $\mu$ where $\mu=1,\cdots,M$. A particle of type $\mu$ hops to its rightmost neighboring lattice site with the rate $p_{\mu}$, provided that the target lattice site is not already occupied, according to the following rule:
$$
\mu \; 0 \; \longrightarrow \; 0 \; \mu \; \mbox{with the rate} \; p_{\mu}.
$$
Note that the particles do not overtake; therefore, the sequence of particles is preserved. On the other hand, the particle's hopping rate is a constant and depends only on the particle type $\mu$.   Let us denote the number of empty lattice sites in front of a particle of type $\mu$ as $n_{\mu}$. In the steady-state the probability of finding the system in the configuration $\{n_{1},\cdots,n_{M}\}$ is given by:
\begin{equation}
\label{PDF1}
P(\{ n_{1},\cdots,n_{M}\})=\frac{1}{Z_{N,M}}Tr(D_{1}E^{n_{1}} \cdots D_{M}E^{n_{M}})
\end{equation}
in which the normalization factor $Z_{N,M}$ is called the canonical partition function of the system. One should also note that $\sum_{\mu=1}^{M}n_{\mu}=N-M$. In (\ref{PDF1}) the matrices $D_{\mu}$ and $E$ stand for the presence of a particle of type $\mu$ and an empty lattice site respectively. It has been shown that these matrices satisfy the following quadratic algebra:
\begin{equation}
\label{Algebra1}
D_{\mu}E=\frac{1}{p_{\mu}}D_{\mu} \;\;,\;\; \mu=1,\cdots,M.
\end{equation}
In \cite{EH05} the authors have introduced a matrix representation for the quadratic algebra (\ref{Algebra1}) as follows:
\begin{equation}
\label{Rep1}
D_{\mu}=\sum_{i=0}^{\infty}f_{\mu}(i)\vert 0 \rangle \langle i \vert \;\; , \;\; E=\sum_{i=0}^{\infty}\vert i+1 \rangle \langle i \vert
\end{equation}
in which $f_{\mu}(i)=\prod_{j=1}^{i}\frac{1}{{p_{\mu}}}=\frac{1}{{p_{\mu}}^i}$. Note that $\{ \vert i \rangle \}$  is the standard basis of an infinite-dimensional vector space that is  $\vert i \rangle_{j}=\delta_{i,j}\;\;\mbox{for}\;\;i,j=0,1,\cdots,\infty$. The steady-state probability distribution (\ref{PDF1}) can now be calculated and one finds:
\begin{equation}
\label{PDF2}
P(\{ n_{1},\cdots,n_{M}\})=\frac{1}{Z_{N,M}}\prod_{\mu=1}^{M}f_{\mu}(n_{\mu})
\end{equation}
in which the normalization factor $Z_{N,M}$ is given by:
\begin{equation}
Z_{N,M} =
 \sum_{n_1,n_2 \ldots n_M}
\delta_{  \sum_{\mu} n_\mu,\ (N-M)}
 \prod_{\mu=1}^{M} f_\mu( n_\mu )
\end{equation}
Let us now consider a generalized disordered driven-diffusive system consisting of different types of hard-core particles. We assume that a particle of type $\mu$ hops one lattice site forward according to the following rule:
$$
\mu \; \underbrace{0 \; \cdots \;0}_{n_{\mu}} \; \mu' \longrightarrow \; 
0 \; \mu \; \underbrace{0 \; \cdots 0}_{n_{\mu}-1} \; \mu' \; \mbox{with the rate} \; u_{\mu}(n_{\mu}).
$$
This is a disordered driven-diffusive system with long-range interactions since the hopping rate of the particle of type $\mu$, that is $u_{\mu}(n_{\mu})$, depends on the number of empty lattice sites in front of it. The steady-state probability distribution of this generalized disordered system $P(\{ n_{1},\cdots,n_{M}\})$ is a product measure and can still be written as a trace of a matrix product similar to (\ref{PDF1}).  It is not surprising since this generalized disordered driven-diffusive system can be mapped onto a ZRP.  Therefore it has a product measure steady-state. In order to verify this, note that the matrices $D_{\mu}$ and $E$ in (\ref{PDF1}) satisfy the following algebra:
\begin{equation}
\label{Algebra2}
D_{\mu}E^{n_{\mu}}D_{{\mu}'}=f_{\mu}(n_{\mu})D_{{\mu}'} \;\; \mbox{for}\;\;\mu,\mu'=1,\cdots,M
\end{equation}
provided that we define $f_{\mu}(i)=\prod_{j=1}^{i}\frac{1}{{u_{\mu}}(j)}$ and that $f_{\mu}(0)=1$. It is easy to verify that the matrix representation (\ref{Rep1}) satisfies the algebra (\ref{Algebra2}). Using the algebra (\ref{Algebra2}), straightforward  calculations show that the steady-state probability distribution of the generalized disordered driven-diffusive system (\ref{PDF1}) can be simplified to (\ref{PDF2}).\\
It should be noted that for the special case $u_{\mu}(n_{\mu})=p_{\mu}$ (equivalently $f_{\mu}(n_{\mu})=\frac{1}{{p_{\mu}}^{n_{\mu}}}$) the matrix representation (\ref{Rep1}) satisfies both (\ref{Algebra1}) and (\ref{Algebra2}); however, for a general $u_{\mu}(n_{\mu})$ the matrix representation (\ref{Rep1}) only satisfies (\ref{Algebra2}).
\section{The heterogeneous zero-range process}
\begin{figure}
\includegraphics[width=3in]{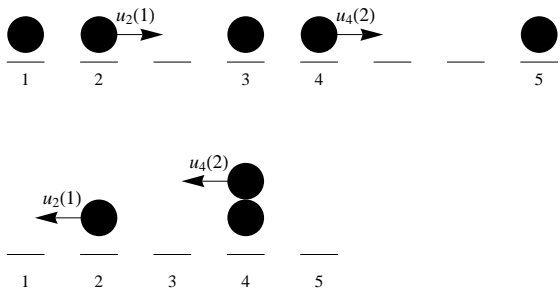}
\caption{\label{Graphics1} Mapping between a disordered driven-diffusive system (above) and a heterogeneous zero-range process (below) with a periodic boundary condition.}
\end{figure}

The generalized disordered driven-diffusive system introduced in the previous section can be mapped onto a HZRP.  As for this mapping a particle of type $\mu$ in the driven-diffusive system should be considered as the $\mu$'th box in the ZRP. The number of empty lattice sites in front of that particle in the driven-diffusive system is then equal to the number of particles in the box associated with that particle. In this case if the number of particles in the $\mu$'th box is $n_{\mu}$ then the rate at which a particle moves from $\mu$'th box to $(\mu-1)$'th box is given by $u_{\mu}(n_{\mu})$. Note that if the particles in the driven-diffusive system move forward, they move backward in the ZRP (see FIG. \ref{Graphics1}) . It can be seen the particle's hopping rate from one box to the next box depends on both the box number and the number of particles inside it, which can be at most of order of the total number of empty lattice sites in the driven-diffusive system. In FIG. \ref{Graphics1} we have provided a simple sketch of such mapping.
The steady-state of the HZRP defined above is still given by (\ref{PDF2}). This has already been obtained in \cite{EH05} using a different approach.\\
In \cite{BM10} the authors have studied a homogeneous ZRP and calculated its steady-state probability distribution using a matrix product method similar to that which we have done here. Their algebra and its matrix representation are special cases of (\ref{Algebra2}) and (\ref{Rep1}). The main difference is that in a homogeneous system the particles' hopping rates do not depend on the box number.
\section{The walk model}
\begin{figure}
\includegraphics[width=2in]{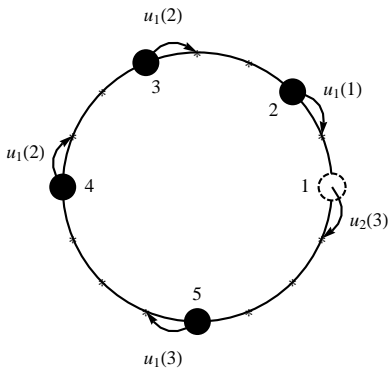}
\caption{\label{Graphics2} A disordered driven-diffusive system on a ring of length $N=16$ containing $M=5$ particles. The first particle (white disk) is considered as a second-class particle while the rest of the particles (black disks) are considered as first-class particles.}
\end{figure}

In \cite{J11} the author has shown that a simple disordered driven-diffusive system, consisting of a finite number of ordinary particles in the presence of a single impurity, which can be mapped onto a HZRP,  has a walk model interpretation. In this section we aim to show that the generalized  HZRP introduced in the previous section has a simple equivalent walk model in a special case.  Note that if the particles' hopping rates are constant and do not depend on the number of empty lattice sites in front of each particle, the model converges into the process introduced and discussed in \cite{E96} and \cite{E00}.\\
Let us start with the definition of the disordered driven-diffusive system. We assume that the driven-diffusive system is defined on a ring of length $N$ and consists of two different types of particles. There is one second-class particle (let us consider it as the first particle) which hops to its rightmost vacant neighboring lattice site with the rate  $u_{2}(n)$ provided that there are $n$ empty lattice sites in front of it. There are also $M-1$ first-class particles on the ring which are considered as $2$nd, $3$rd, $\cdots$ and $M$th particles. Each first-class particle hops to its rightmost vacant neighboring lattice site with the rate  $u_{1}(n)$ provided that there are $n$ empty lattice sites in front of it. A simple example is plotted in FIG. \ref{Graphics2}. We associate the operators $E$, $D_{1}$ and $D_{2}$ to an empty lattice site, a first-class particle and a second-class particle respectively.  These operators satisfy the algebra (\ref{Algebra2}) for $\mu=1,2$. Their matrix representation is also given by (\ref{Rep1}) in which $f_{\mu}(i)=\prod_{j=1}^{i}\frac{1}{{u_{\mu}}(j)} $ for $\mu=1,2$. It is much easier to work in a grand-canonical ensemble in which the number of first-class particles is controlled by a fugacity. According to the matrix product method discussed above and following \cite{J11} the grand-canonical partition function of this system can be written as:
\begin{equation}
\label{GCPF1}
Z_{N}(z)=Tr(C^{N-1}D_{2})
\end{equation}
in which $C=zD_{1}+E$. Now one can rewrite (\ref{GCPF1}) using the matrix representation of $D_{2}$ given in (\ref{Rep1}) as:
\begin{equation}
\label{GCPF2}
Z_{N}(z)=\sum_{j=0}^{N-1}f_{2}(j)\langle j \vert C^{N-1}\vert 0 \rangle
\end{equation}
in which we have used the fact that:
\begin{equation}
\label{Rule}
C\vert j\rangle= zf_{1}(j)\vert 0 \rangle + \vert j+1 \rangle.
\end{equation}
Note that $C^{N-1}$ on $\vert 0 \rangle$ can be written as a linear superposition of $\{ \vert 0 \rangle,\cdots,\vert N-1 \rangle \}$. Expanding (\ref{GCPF2}) as a power series in $z$, the canonical partition function of the generalized driven-diffusive system is equal to the coefficient of $z^{M-1}$. \\
This disordered driven-diffusive system can be mapped onto a HZRP consisting of $M$ boxes and $N-M$ similar particles.  The boxes are divided into two different types. The rate at which a particle leaves the first box $\mu=1$ is given by $u_{1}(n_{1})$ while the rate at which a particle leaves the other boxes $\mu=2,\cdots,M$ is given by $u_{2}(n_{\mu})$. The mapping guaranties that the partition function of the disordered driven-diffusive system is equal to that of the HZRP. \\
In \cite{J11} it has been shown that a much simpler HZRP in which the particles leave the boxes at a constant rate can be mapped onto a two-dimensional walk model. In that which follows we will show that the HZRP defined above can also be mapped onto a walk model.
One can see that the expression (\ref{GCPF2}) can be interpreted as the partition function of a simple random walk model if we consider $C$ as the transfer matrix of a two-dimensional random walk defined on the lattice sites of a two-dimensional plane $\mathbb{Z}^2_{+}=\{(i,j):i , j\ge 0 \mbox{\; are integers} \}$. Each step taken by the random walker has a weight. The random walker moves according to the following rules: 
\begin{itemize}
\item{ For $i \geq j$ from the lattice site $(i,j)$ to $(i+1,j+1)$ with a weight $1$ where $i,j=0,1,2,\cdots,\infty$. This is called an upward step.}
\item{ For $i \geq j$ from the lattice site $(i,j)$ to $(i+1,0)$ with a weight $zf_{1}(j)$ where $i,j=0,1,2,\cdots,\infty$. This is called a horizontal (for $j=0$) or a downward (for $j\neq0$) step.}
\end{itemize}
Keeping in mind that the random walker does not take any steps in the negative $i$-direction, we assume that it starts from the origin $(0,0)$ and takes $N-1$ consecutive steps according to the above mentioned rules. Finally the random walker should be at the lattice site $(N-1,j)$ where $j=0,\cdots,N-1$. We define the weight of a path of length $N-1$ as a product of the weights of the steps taken by the random walker times an extra weight associated with the destination point that is $f_{2}(j)$ if the random walker's destination lattice site is $(N-1,j)$. It is now easy to see why the transfer matrix of the walk model is exactly equal to $C$ defined in (\ref{GCPF1}) and (\ref{Rule}). The partition function of the walk model is also given by (\ref{GCPF2}).
\section{The canonical partition function}
The disordered driven-diffusive system defined in previous section contains exactly $M$ particles. When we map this system onto a HZRP it contains $M$ boxes. As we saw the grand-canonical partition function of the disordered driven-diffusive system (or equivalently the HZRP) is equal to that of the walk model defined by the rules in previous section. It is easy to verify that if we are interested in the canonical partition function of the disordered driven-diffusive system, we need the sum of the weights of the paths which contain exactly $N-M$ upward steps or equivalently $M-1$ downward and horizontal steps.  This is because each downward or horizontal step contains one $z$ in the weight of that step. 
In what follows we show how we can calculate the canonical partition function of the disordered driven-diffusive system.\\
The expression (\ref{GCPF2}) can be rewritten as:
\begin{equation}
\label{GCPF3}
Z_{N}(z)=\sum_{k_{0}=0}^{N-1}f_{2}(N-k_{0}-1) F(k_{0})
\end{equation}
in which we have defined $F(k_{0})=\langle 0 \vert C^{k_{0}} \vert 0 \rangle$. Clearly we have $F(0)=1$. Using the matrix representation (\ref{Rep1}) one finds that $F(k_{0})$ satisfies the following recursion relation:
\begin{equation}
\label{RR}
F(k_{0})=\sum_{k_{1}=0}^{k_{0}-1}zf_{1}(k_{1})F(k_{0}-k_{1}-1)
\end{equation}
for $k_{0}=1,2,\cdots,N-1$. After some calculations the recursion relation (\ref{RR}) can be solved and one finds:
\begin{equation}
\label{RRSolv}
F(k_{0})=\sum_{i=0}^{k_{0}-1}[\prod_{j=1}^{i}{\big (} \sum_{k_{j}=1}^{k_{j-1}-1}zf_{1}(k_{j-1}-k_{j}-1){ \big )}] zf_{1}(k_{i}-1).
\end{equation}
Care should be taken in using (\ref{RRSolv}) in this compact form since its first term for $i=0$ is defined to be $zf_{1}(k_{0}-1)$. As we mentioned above, we are interested in the canonical partition function that is the coefficient of $z^{M-1}$ in (\ref{GCPF3}).  This can be achieved using (\ref{GCPF3}) and (\ref{RRSolv}) and the result is:
\begin{widetext}
\begin{equation}
\label{CPF1}
Z_{N,M}=\sum_{k_{0}=1}^{N-1}[\prod_{j=1}^{M-2}{\big (} \sum_{k_{j}=1}^{k_{j-1}-1}f_{1}(k_{j-1}-k_{j}-1){ \big )}]f_{1}(k_{M-2}-1)f_{2}(N-k_{0}-1).
\end{equation}
\end{widetext}
This is the canonical partition function of three different models which can be mapped onto each other; a disordered driven-diffusive system, a HZRP and a two-dimensional walk model as described in previous sections. Let us investigate two special cases studied in \cite{E96} and \cite{AEM04}. In the first case we have $u_{1}(n)=1$ and $u_{2}(n)=p$ (equivalently $f_{1}(n)=1$ and $f_{2}(n)=\frac{1}{p^{n}}$). Using the fact that:
\begin{equation}
\sum_{k_{1}=1}^{k_{0}-1}\sum_{k_{2}=1}^{k_{1}-1}\cdots\sum_{k_{M-2}=1}^{k_{M-3}-1}1=
\left( \begin{array}{c}
k_{0}-1\\
M-2
\end{array} \right)
\end{equation}
in which 
$\left( \begin{array}{c}
a\\
b
\end{array} \right)=\frac{a!}{b!(a-b)!}$ is the usual binomial coefficient, it is easy to see that (\ref{CPF1}) results in:
\begin{equation}
Z_{N,M}=\sum_{k_{0}=0}^{N-M}
\left( \begin{array}{c}
N-k_{0}-2\\
M-2
\end{array} \right) p^{-k_{0}}
\end{equation}
which is exactly the canonical partition function of the system studied in \cite{E96}. In the second case introduced and studied in \cite{AEM04} we have $u_{1}(n)=1$ and $u_{2}(n)=p(1+\frac{\lambda}{n})$. For $\lambda=0$ it converges to the first case mentioned before. It is easy to verify that $f_{1}(n)=1$ and $f_{2}(n)=\frac{n!}{p^{n}(1+\lambda)_{n}}$ in which $(a)_{n}$ is the Pochhammer symbol defined as:
\begin{equation}
(a)_{n}=a(a+1)(a+2)\cdots(a+n-1)\;\;,\;\;(a)_{0}=1.
\end{equation} 
Replacing these into (\ref{CPF1}) one finds:
\begin{equation}
Z_{N,M}=\sum_{k_{0}=0}^{N-M}
\left( \begin{array}{c}
N-k_{0}-2\\
M-2
\end{array} \right) \frac{k_{0}!}{p^{k_{0}}(1+\lambda)_{k_{0}}}.
\end{equation}
In terms of the number of empty lattice sites, that is $L:=N-M$, one recovers the expression $Z_{L,M}$ which has already been obtained in \cite{AEM04}.
\section{Concluding remarks}
In this paper we have investigated a disordered driven-diffusive system defined on a ring which can be mapped onto a HZRP. We have shown that the grand-canonical partition function of the HZRP can be interpreted as the partition function of a two-dimensional walk model. We have obtained exact expressions for both the grand-canonical and canonical partition functions of the HZRP. We have presented two simple examples which confirm the correctness of our calculations. Apart from the special driven-diffusive system studied in this paper, in which a single second-class particle beside a finite number of first-class particles, lie on the lattice and hop stochastically in  a preferred direction, one can also consider other cases. For instance, a disordered driven-diffusive system in which the backward hopping of the particles is also included similar to the one studied in \cite{E96}. In terms of the walk model this results in horizontal movements at all heights. In \cite{EH05} a variety of the zero-rage processes have been investigated. It would be interesting to investigate the walk models associated with different types of these processes.

\end{document}